\documentclass{ICEAA-IEEE_APWC}
\pdfoutput=1
\usepackage[numbers,sort]{natbib}
\usepackage{amsmath}
\usepackage{booktabs}
\usepackage{url}

\usepackage{tikz}
\usepackage{textcomp}
\usepackage{hyperref}
\usepackage{lipsum}

\newcommand\copyrighttext{%
  \footnotesize \textcopyright 2015 IEEE. Personal use of this material is permitted.
  Permission from IEEE must be obtained for all other uses, in any current or future 
  media, including reprinting/republishing this material for advertising or promotional 
  purposes, creating new collective works, for resale or redistribution to servers or 
  lists, or reuse of any copyrighted component of this work in other works. }
\newcommand\copyrightnotice{%
\begin{tikzpicture}[remember picture,overlay]
\node[anchor=south,yshift=10pt] at (current page.south) {\fbox{\parbox{\dimexpr\textwidth-\fboxsep-\fboxrule\relax}{\copyrighttext}}};
\end{tikzpicture}%
}

\makeatletter
\let\@fnsymbol\@arabic
\makeatother

\title{Measured Sensitivity of the First Mark II Phased Array Feed on an ASKAP Antenna}

\author{%
A. P. Chippendale\thanks{CSIRO Astronomy and Space Science, PO Box 76, Epping, NSW 1710, Australia, e-mail: \texttt{Aaron.Chippendale@csiro.au}, tel.: +61 2 9372 4296.}  \and A. J. Brown$^1$ \and R. J. Beresford$^1$ \and G. A. Hampson$^1$ \and A. Macleod$^1$  \and R. D. Shaw$^1$ \and M. L. Brothers$^1$ \and C. Cantrall$^1$ \and A. R. Forsyth$^1$ \and S. G. Hay\thanks{CSIRO Digital Productivity, PO Box 76, Epping, NSW 1710, Australia, e-mail: \texttt{Stuart.Hay@csiro.au}} \and M. Leach$^1$}

\begin{document}
\maketitle
\copyrightnotice

\begin{abstract}
This paper presents the measured sensitivity of CSIRO's first Mk.~II phased array feed (PAF) on an ASKAP antenna.  The Mk.~II achieves a minimum system-temperature-over-efficiency $T_\mathrm{sys}/\eta$ of 78~K at 1.23~GHz and is 95~K or better from 835~MHz to 1.8~GHz.  This PAF was designed for the Australian SKA Pathfinder telescope to demonstrate fast astronomical surveys with a wide field of view for the Square Kilometre Array (SKA).  
\end{abstract}

\section{INTRODUCTION}

\begin{figure}
\centering
\includegraphics[width=0.8\columnwidth]{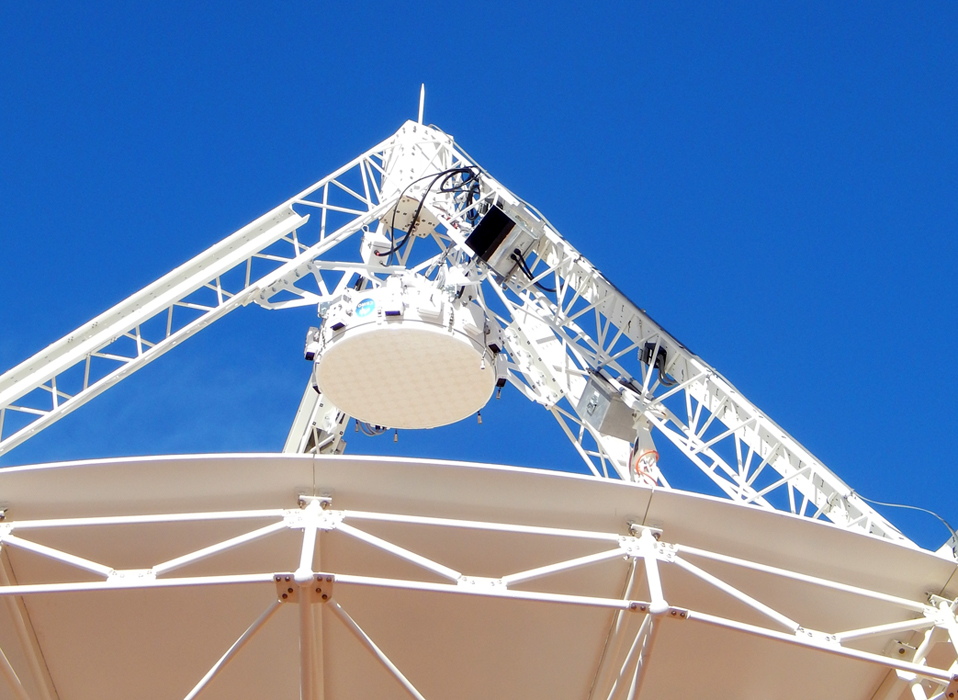}
\caption[]{The first Mk.~II ASKAP PAF installed on antenna 29 at the MRO.}
\label{fig:ant29}
\end{figure}

We present preliminary measurements of the sensitivity of CSIRO's first Mk.~II phased array feed (PAF) \cite{Hampson2012} on an ASKAP antenna as shown in Figure \ref{fig:ant29}.  Over the next two years, Mk.~II PAFs will be installed on thirty 12~m parabolic reflectors of the Australian SKA Pathfinder telescope (ASKAP) \cite{DeBoer2009} to demonstrate fast astronomical surveys with a wide field of view for the Square Kilometre Array (SKA).  The SKA is an international project to build the world's largest radio telescope with a square kilometre of collecting area  \citep{Dewdney2009}.

\begin{figure*}[t!]
\centering
\includegraphics[width=0.7\textwidth]{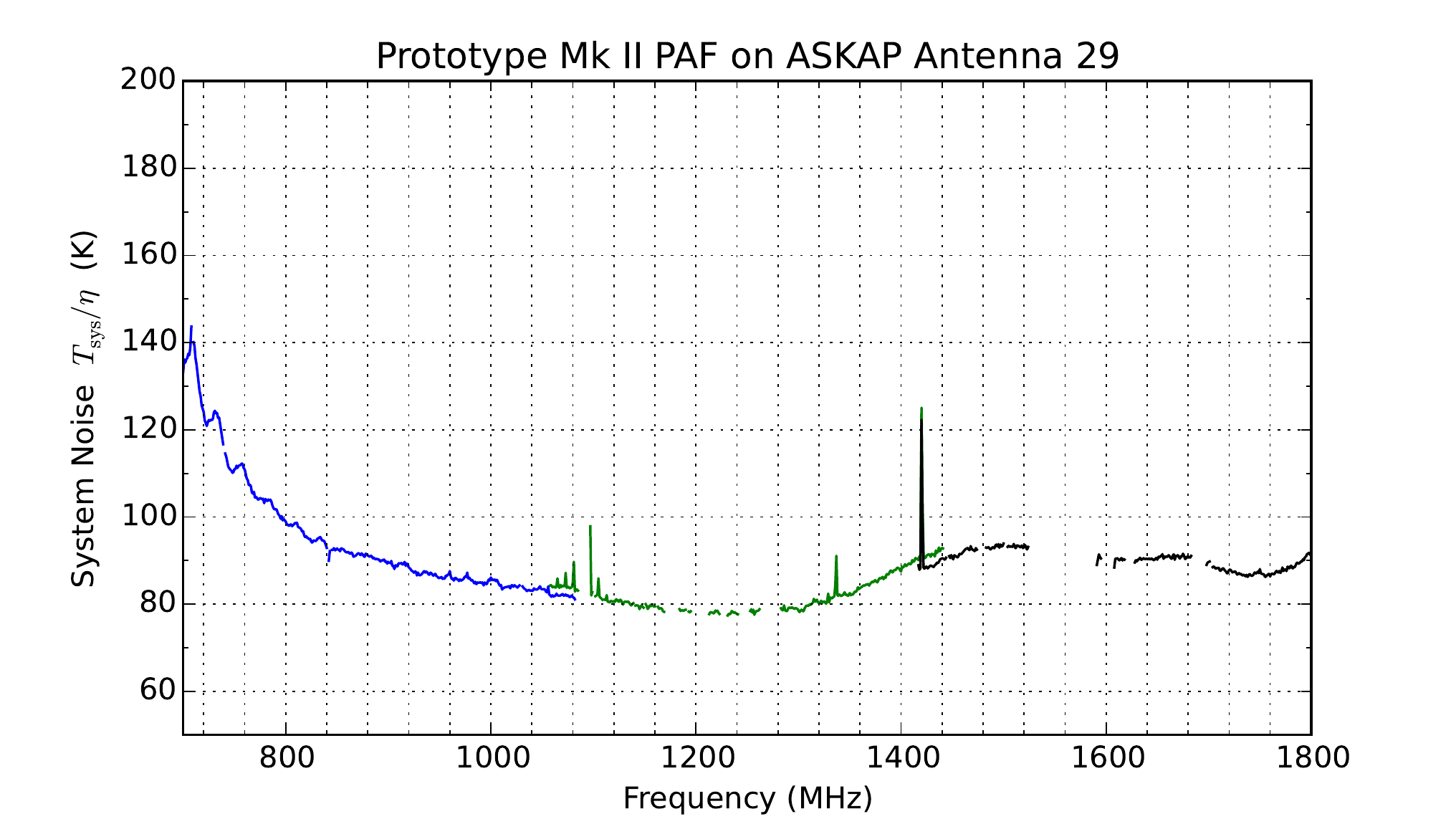}
\caption[]{Preliminary beam equivalent system-temperature-over-efficiency $T_\mathrm{sys}/\eta$ for the first Mk.~II PAF on an ASKAP antenna.  This is for a single-polarization boresight beam with maximum signal-to-noise ratio weights.  Beamforming and sensitivity measurements were made using Tau A with 10~s integration for both on and off-source positions.  Line colours show different digital receiver bands.  Gaps in frequency coverage are due to radio-frequency interference (RFI) drowning out Tau A in the beamforming process.  Uncertainty in $T_\mathrm{sys}/\eta$ is at least 6\% due to uncertainty in the flux of Tau A.}
\label{fig:ak29tsyseta}
\end{figure*}

Figure \ref{fig:ak29tsyseta} shows that the Mk.~II achieves a minimum system-temperature-over-efficiency $T_\mathrm{sys}/\eta$ of 78~K at 1.23~GHz and is 95~K or better from 835~MHz to 1.8~GHz with the receiver near room temperature.  By comparison, the CSIRO Mk.~I PAF is 95~K or better only from 735~MHz to 1.2~GHz \citep{Hotan2014}. This significant improvement was achieved via enhanced antenna array and low-noise amplifier (LNA) designs \citep{Shaw2012}.  Both Mk.~I and Mk.~II ASKAP PAFs are based on a connected-element ``chequerboard'' array \citep{Hay2008} that is dual-polarized, low-profile, and inherently wide-band.  

The only other PAF being built in comparable numbers to the ASKAP PAFs is ASTRON's APERTIF PAF for 12 antennas of the Westerbork Synthesis Radio Telescope.  It is specified for $T_\mathrm{sys}/\eta=93$~K over 1.13~GHz to 1.75~GHz \citep{Cappellen2014}.  Next steps to improve PAF sensitivity include the adoption of lower-noise transistors \citep{Shaw2015} and cryogenically cooling PAFs as recently demonstrated on the Green Bank Telescope \citep{Roshi2015}.  

\section{MEASUREMENT SYSTEM}
Figure \ref{fig:ant29} shows the prototype Mk.~II PAF installed on ASKAP antenna 29 at the Murchison Radio-astronomy Observatory (MRO) in Western Australia.  Antenna 29 is cabled so a receiver can be tested as a PAF at the antenna's focus or as an aperture array on the ground nearby.

Radio-frequency (RF) signals from each of the 188 ``chequerboard'' ports are modulated onto optical fibre links within the PAF package.  The RF signals then travel over 1.4~km of optical fibre to the digital receiver in the ASKAP control building.

The ASKAP digital receiver \citep{Brown2014} directly samples 192 signals (188 from the PAF and 4 spare) and then uses an oversampled poly-phase filter bank \citep{Tuthill2012} to divide these signals into 1~MHz channels.  Each directly-sampled band has approximately 600~MHz RF bandwidth, but only 384~MHz are passed to the beamformer in the current configuration.  A particular contiguous 384~MHz sub-band is selected for beamforming via a software command to set the centre frequency.    

ASKAP's beamformer \citep{Hampson2014} includes a firmware Array Covariance Matrix (ACM) module to calculate the receiver-output sample covariance matrix
\begin{equation}
  \mathbf{R} = \frac{1}{L}\sum_{n=1}^{L}\mathbf{x}(n)\mathbf{x}^{H}(n)
  \label{eq:scm}
\end{equation}
where $\mathbf{x}(n)$ is the $n^{\text{th}}$ time sample of the column vector of 192 complex array-port voltages $\mathbf{x}(t)$ for a single 1~MHz coarse filter bank (CFB) channel.  Rather than using the online firmware beamformer, we downloaded the ACM output and calculated beamformed power $P$ for weight vector $\mathbf{w}$ by
\begin{equation}
  P = \mathbf{w}^H\mathbf{R}\mathbf{w}.
  \label{eq:bfpower}
\end{equation}   
This allows offline experimentation with arbitrary beamformer weights.
\section{SENSITIVITY MEASUREMENTS}
We deduce the on-dish noise performance of the ASKAP PAF from the Y-factor power ratio between measurements of a standard astronomical source and nearby empty sky \citep{Hayman2010, Landon2010}.  The Y-factor and therefore the telescope sensitivity are a function of the beamformer weights
\begin{equation}
  Y = \frac{P_{\mathrm{on}}}{P_{\mathrm{off}}} = \frac{\mathbf{w}^H\mathbf{R}_{\mathrm{on}}\mathbf{w}}{\mathbf{w}^H\mathbf{R}_{\mathrm{off}}\mathbf{w}}.
  \label{eq:yfact}
\end{equation}
Here we measure beamformed power $P_\mathrm{on}$ towards an astronomical source of known flux and $P_\mathrm{off}$ towards nearby empty sky.  

\begin{table*}[t!]
\begin{center}
\footnotesize
\begin{tabular}{ccccccc}
\toprule
  Band & Sampled & Digtal Centre         & Beamformer  & Observation & \multicolumn{2}{c}{Elevation}\\ 
       & RF Band & Frequency & Band & UTC Epoch & Tau A & off-source \\
       &  (MHz)          & (MHz) & (MHz)& (yyyymmddhhmmss) & (deg) & (deg) \\ \midrule
  1 & 700-1200 & 891 &   700-1083 & 20140909210010 &  37.2 & 34.5\\
  2 & 840-1440 & 1249 & 1058-1441 & 20140911203019 &  34.5 & 31.7\\
  3 & 1400-1800 & 1608 & 1417-1800 & 20140911215449 & 41.0 & 40.0\\
  4 & 600-700  & 575 &   384-767 & \multicolumn{3}{c}{ \rule[1mm]{12mm}{.1pt} measurement pending \rule[1mm]{12mm}{.4pt}} \\
  \bottomrule
\end{tabular}
\end{center}
\caption{Measurement parameters.}\label{tab:measparams}
\end{table*}

The ratio of system temperature $T_\mathrm{sys}$ to antenna efficiency $\eta$ is then given by \citep{Hayman2010}  
\begin{equation}
\frac{T_\mathrm{sys}}{\eta} = \frac{AS}{2k_\mathrm{B}(Y-1)}
\label{eq:tsys}
\end{equation} 
where $A$ is the geometric antenna (reflector) area, $S$ is the known flux of the reference source and $k_\mathrm{B}$ is Boltzmann's constant.  Antenna efficiency is defined by $\eta=\eta_\mathrm{ap}\eta_\mathrm{rad}$, the product of aperture efficiency and radiation efficiency.  The efficiencies and system noise temperature of receiving arrays are more fully defined in \citep{Warnick2010}.  They both depend on the beamformer weights as can be implied from equations \eqref{eq:yfact} and \eqref{eq:tsys}.

Equation \eqref{eq:tsys} shows that $T_\mathrm{sys}/\eta$ is minimised by maximising $Y$.  Rearranging \eqref{eq:yfact} reveals an eigenvalue problem 
\begin{equation}
  \mathbf{R}_{\mathrm{off}}^{-1}\mathbf{R}_{\mathrm{on}}\mathbf{w} = Y\mathbf{w}.
  \label{eq:evalprob}
\end{equation}
Maximum beamformed Y-factor corresponds to the dominant eigenvalue of $\mathbf{R}_\mathrm{off}^{-1} \mathbf{R}_\mathrm{on}$ and is achieved with beamformer weights set to the dominant eigenvector \citep{Landon2010}.  In this work we included all ports of both polarizations in the beamformer weight solution by using the full $188\times188$ matrix $\mathbf{R}$ when solving equation \eqref{eq:evalprob}.  Figure \ref{fig:ak29tsyseta} shows the resulting sensitivity for the beam corresponding to the strongest of two dominant eigenvalues.  This should be the overall best-case maximum sensitivity beam, but its polarization will be  biased by any polarization in the signal field (calibrator) or noise fields (eg. spillover and receiver noise).  

Table \ref{tab:measparams} summarises the measurements made for this work.  On and off-source measurements were made in three of ASKAP's RF bands over two days in September 2014.  The off-source position was a $+7^\circ$ offset in right ascension.  We used Taurus A as the flux reference assuming the flux model of \citep{Baars1977}.  Tau A is 0.5\% and 3.5\% polarized at wavelengths of 20~cm and 30~cm respectively \citep{Kuzmin1966}.

Care should be taken when comparing our measurements to those from the northern hemisphere using Cas A as the flux reference.  The strength of Cas A is evolving non-uniformly with time \citep{Vinyajkin1997}.  This necessitates ongoing monitoring of flux calibration sources concurrent with their use to measure the sensitivity of new telescope receivers \citep{Kraus2015}.


Single-dish beamforming with a PAF on a 12~m reflector antenna in the southern hemisphere is challenging as Tau A, the strongest available unresolved source, has a flux of 980~Jy at 1.25~GHz.  This is weaker than the system equivalent flux density of 2,000~Jy at the same frequency.  Most single-dish beamforming for the Mk.~I BETA PAFs has been performed using the Sun, which although slightly resolved by a 12~m reflector antenna, provides a much higher signal-to-noise ratio ($S/N$) of order 100 to 1,000 \citep{Hotan2014}.  In this work we have for the first time successfully beamformed on a single 12~m ASKAP antenna with an instantaneous $S/N$ less than unity.  This is important for commissioning, but we expect the full ASKAP to make high $S/N$ beamforming solutions via interferometry \citep{Chippendale2010, Hayman2010}. 

We also measured aperture-array sensitivity via the method of \citep{Chippendale2014} and found a beam equivalent system noise temperature referenced to the sky of $\hat{T}_\mathrm{sys}<50$~K from 800~MHz to 1.7~GHz.  Beamformer weights and spillover differ between aperture array measurement of $\hat{T}_\mathrm{sys}=T_\mathrm{sys}/\eta_\mathrm{rad}$ and on-dish measurement of $T_\mathrm{sys}/\eta$.  Measuring efficiency is not as simple as dividing one by the other.

\section{CONCLUSION}
Our $T_\mathrm{sys}/\eta$ measurement for the first Mk.~II PAF shows a doubling of low-noise bandwidth over the Mk.~I PAF results in \cite{Hotan2014}.  Above 1.4~GHz, sensitivity has been doubled and survey speed should be quadrupled.  The Mk.~II PAF has operated reliably at the MRO and near continuously since it was installed on antenna 29 in September 2014. This has included numerous summer weeks with maximum daily site temperatures regularly exceeding 40$^\circ\text{C}$.



\section*{Acknowledgements}
The Australian SKA Pathfinder is part of the Australia Telescope National Facility which is funded by the Commonwealth of Australia for operation as a National Facility managed by CSIRO. This scientific work uses data obtained from the Murchison Radio-astronomy Observatory (MRO), which is jointly funded by the Commonwealth Government of Australia and State Government of Western Australia. The MRO is managed by the CSIRO, who also provide operational support to ASKAP. We acknowledge the Wajarri Yamatji people as the traditional owners of the Observatory site.

Creating the Mk.~II ASKAP PAF was a large team effort conceived and lead by Grant Hampson.  Project management included Adam Macleod, Adrian Rispler and Ant Schinckel.  John Bunton provided key concepts and critical specifications across the system as ASKAP project engineer.   

The  prototype team included Aaron Chippendale (leader), Ron Beresford, Andrew Brown, Steve Broadhurst, Michael Brothers, Chris Cantrall, Warren Chandler, David Chandler, Paul Doherty, Ross Forsyth, Dezso Kiraly, Jeganathan Kanapathippillai, Mark Leach, Neale Morison and Paul Roberts.  The firmware team included Andrew Brown (leader), Stephan Neuhold (manager), John Tuthill, Tim Bateman, Craig Haskins and John Bunton.  The production team included Steve Barker (leader), Matthew Shields (project engineer), Alan Ng, Andrew Ng, Aaron Sanders, Raji Chekkala, Dezso Kiraly, Wan Cheng and Neale Morison.  Electromagnetic designers and validators included Stuart Hay, Robert Shaw, Doug Hayman and Russell Gough.  

The Marsfield Workshop supported the prototype work, in particular Paul Cooper, Michael Bourne, Minh Hyunh, Raymond Moncay and Michael Death.  Additional assistance was given to the prototype work by Santiago Castillo, Yoon Chung, Daniel Gain, Li Li, Simon Mackay and Les Reilly.  Deployment to the MRO and on-site testing was supported by Suzy Jackson, Michael Reay, John Morris, Lou Puls, Ryan McConigley, Shaun Amy, Aidan Hotan and John Reynolds.  

\bibliographystyle{unsrtnat}
\bibliography{IEEEabrv,apj-jour,dish_pdr_ade}


\end{document}